
%
%
%
\input amstex  
\documentstyle{amsppt}
\NoBlackBoxes
\magnification \magstep 1

\def\today{\number\day\space\ifcase\month\or January\or February\or
March\or April\or May\or June\or July\or August\or September\or
October\or November\or December\fi\space\number\year}

\def\image{\operatorname{Image}}
\def\tr{\operatorname{tr}}
\def\identity{\operatorname{identity}}
\def\End{\operatorname{End}}
\def\surgery{\operatorname{surgery}}

\def\finitedimensional{fi\-nite-di\-men\-sion\-al} 

\def\bra{\langle}    
\def\C {|}
\def\ket{\rangle}
\def\NOT{{\smc not }} 
\def\bbar{{\overline b}}
\define\SU{\operatorname{SU}}
\define\Sl{\operatorname{sl}}

\newcount\refc \global\refc=0   
\def\refer#1{\cite{\reference{#1}}}

\def\reference#1{\global\advance\refc by 1{\number\refc\relax}}

\def\refname#1{\xdef#1{\number\refc}}

\topmatter
\title Quantum Gravity as Topological Quantum Field Theory
\endtitle
\author John W. Barrett\endauthor
\date 18 June 1995\enddate
\address
Department of Mathematics,
University of Nottingham,
University Park,
Nottingham,
NG7 2RD, UK
\endaddress
\email jwb\@maths.nott.ac.uk \endemail
\thanks PACS 04.60.-m\endthanks    
\abstract
The physics of quantum gravity is discussed within the framework of topological
quantum field theory. Some of the principles are illustrated with examples
taken from theories in which space-time is three dimensional.
\endabstract
\toc
\widestnumber\head{III}
\head I. Introduction \endhead
\head II. A Combinatorial Problem \endhead
\head III. The Space of Quantum States\endhead
\head IV. Global Observables \endhead
\head V. Examples of TQFTs in Dimension Three \endhead
\head VI. Local Observables \endhead
\endtoc
\endtopmatter

\document

\head{I. Introduction}\endhead

{}From the point of view of physics, a quantum theory of gravity is not a
precisely
defined concept. There are two limits, $G\to 0$ and $\hbar\to 0$, which
give particle physics and general relativity, both of which are well
understood.
Apart from this, there are few constraints. It is therefore necessary to make
a series of assumptions of a general nature. Thus from a mathematical point
of view, one would like to keep the general features of quantum theory
on the one hand and the topological framework of space-time on the other.
Both of the features are embraced by the precise notion of a topological
quantum field theory. The aim of this paper is to argue the thesis that
the notion of a topological quantum field theory is sufficiently broad
and general that it is a proper framework for quantum gravity, and that it
encompasses many of the traditional approaches to the subject.

In this theory, the manifold is treated as an external, or unquantised
object, whereas the metric tensor is certainly regarded as quantum mechanical.
An intuitive picture of this is given by the functional integral
approach to quantum gravity, where the idea is to integrate over the
space of all metrics on a given manifold, subject to certain constraints
associated with some given observables. In fact, one can think of the axioms
of TQFT as a way of encoding the desired properties of this functional
 integral.
Continuing with this intuitive picture, one can think of topological objects,
such as manifolds, or particularly chosen submanifolds of a manifold as
labels or names for quantum mechanical observables; the metrics and other
fields being the spectra of these observables.

One of the main features of TQFT is that the way that diffeomorphisms of
manifolds act on the algebraic objects of the theory is stated clearly at the
outset. The consequences of this are surprisingly rich, and include
direct analogues in this theory of the fact that the wavefunctions one can
construct satisfy the equations known as momentum and Hamiltonian
constraints in the canonical quantisation approach.

The most important reason for working with this level of abstraction is
that {\it there are examples}. The most well-known examples are for space-time
dimension three, including theories related to gravity.
However examples exist in all dimensions, but none are known to be interesting,
as far as quantum gravity is concerned, in dimension four. There is no
prima facie reason why quantum gravity in dimension four may not exist within
the framework of TQFT. This is despite the existence of propagating
gravitational waves in four-dimensional general relativity, and the
expectation that propagating gravitons play a role in quantum gravity.
This phenomenon appears to require the use of infinite-dimensional state spaces
or an effective substitute, which has some interesting consequences discussed
in section {\smc V}, which may require some restriction on the topologies
considered. The subject of quantum cosmology is in many ways the forerunner of
the present theory, the desiderata for the functional integral including the
axioms of TQFT.

A lot of work on quantum gravity can be classified in two types. On the
one hand there is the `$\text{particle physics}+\epsilon$' approach, where
one takes models of particle physics and perturbs them by adding new fields,
more dimensions, extended structures or more elaborate quantisation schemes.
In this approach, one hopes that the `correct' model will emerge, and then
it will become apparent what the physical and mathematical principles are
with which to reformulate the theory on to a satisfactory conceptual footing.
On the other hand there are the `fundamentalists', who regard the revision
of the basic notions of space, time, set theory, quantum mechanics or logic as
a necessary preliminary before any proper progress can be made.

The problem with the later approach is that without the guide of some good
examples, there is no reason to prefer any one modification of these basic
notions to another one.  The importance of the three-dimensional
gravity and related TQFTs is that they can be viewed from either the
fundamentalist or the particle physics standpoints. It shows at once that
the mathematical formalism of topological quantum field theory is not empty
and can deliver a theory in which metrics play a role, and on the other hand
allows one to test some cherished assumptions about the nature of observables
and the relationship of classical general relativity to quantum gravity.

\head II. A Combinatorial Problem \endhead

In a topological quantum field theory, the simplest case to consider is a
compact oriented space-time manifold $M$, which is closed (without boundary).
In this case, the
TQFT defines a complex number, $Z(M)$, which is called the partition
function of the manifold, due to the analogy with statistical mechanics.
This number is to be thought of as the result of performing a functional
integral over some fields, such as a metric tensor field, on $M$.

It is natural, and seemingly innocuous, to require that the partition
functions of isomorphic manifolds are the same number. This is called an
invariant of manifolds. This leads directly
to a combinatorial problem to define a partition function for all the
manifolds of a given dimension.
A presentation of a manifold is a recipe for constructing it from a number
of elementary pieces, such as balls or simplexes, by some combinatorial
process which involves a finite amount of information about how the pieces
are assembled. Probably the most well-known method is to present a manifold
by a triangulation. In this method, the pieces are the simplexes and the
information specifies which faces are glued to each other. The combinatorial
problem is to give a computation
of $Z(M)$ from the information specifying the presentation of $M$ in such
a way that the same number is determined for any two different presentations
of the same manifold.

This problem differs sharply in different dimensions. In dimension two,
the closed manifolds are all isomorphic to a sphere with a number of
handles attached. These are readily distinguished when a manifold is
defined by a presentation by calculating the Euler number. Thus, $Z(M)$
is a recursive function of the Euler number.

In dimension three, there is no known list of invariants which specify
a manifold uniquely up to isomorphism, and indeed it is not known even
whether this is possible. Therefore there is no known limit to the
complexity or subtlety of new invariants. The same remarks apply to
simply-connected differentiable manifolds of dimension four. A new feature
in dimension four is that inequivalent differentiable manifolds can be
homeomorphic, for example the (non-compact) space $\Bbb R^4$ has a continuous
family of inequivalent differential structures.

What are the implications of this zoo of differentiable manifolds? The main
point is that the manifolds provide observables for the theory, and so
a rich structure for the observables provides the framework for a non-trivial
quantum theory. In this sense, the complexity of the theory of manifolds
provides a worthy arena for the subject of quantum gravity.

A second point is that the character of these theories will
be very different in different dimensions. Quantum gravity is also necessarily
different in these different dimensions because general relativity is.
General relativity in dimension two is trivial, because the action is just
the Euler number. In dimension three the theory is not trivial, but the
solutions are constant curvature metrics or connections. In dimension four,
there are propagating gravitational waves.

\head III. The Space of Quantum States \endhead
In quantum mechanics, one usually has a space of quantum states associated
to a given physical system. Often this space of states refers to a
particular instant of time, which can be represented in general relativity by
a space-like hypersurface. In topological quantum field theory, this
vector space appears, as part of the definition, when the space-time
manifold $M$ has a boundary. The general framework for a theory of
dimension $d$ is as follows.

Every closed $(d-1)$-dimensional manifold $\Sigma$ has associated to it a
vector
space, $V(\Sigma)$, and every $d$-manifold $M$ determines a vector $Z(M)$
in the state space $V(\partial M)$ which is associated to its boundary.

This accords with the previous use of the notation $Z(M)$ for an invariant
of a closed manifold, for in this case $V(\emptyset)=\Bbb C$,
so $Z(M)\in\Bbb C$ when the boundary of $M$ is empty.

Hawking's no-boundary proposal \refer{}\refname{\Hawking} for the wavefunction
of the universe is an
example of this formalism; there the state space of $S^3$ is defined to
be the vector space of functions on the space of Riemannian metrics on $S^3$,
and the wavefunction of the universe is defined to be the result of
a functional integral on the four-dimensional ball,
$$Z(B^4)\in V(S^3).$$

The data in the TQFT, the state spaces and partition functions, have to satisfy
a number of conditions, which were given by Atiyah \refer{TQFT}\refname\Atiyah.
These
can be summarised as
\roster \item The rules of quantum mechanics for composing amplitudes.
\item Functoriality, or the correct behaviour under diffeomorphisms of
manifolds.
\endroster
The collection of conditions which comes under (1) will become
clear in the following discussion. It includes the condition that for
the disjoint union of two closed $(d-1)$-manifolds $\Sigma_1$ and $\Sigma_2$,
$$V(\Sigma_1\cup \Sigma_2)=V(\Sigma_1)\otimes V(\Sigma_2).$$
This accords with the normal rule in quantum mechanics that the state
space for two systems isolated from each other is the tensor product space.
In this topological theory, the stipulation that the systems are isolated
from each other has been translated into the topological condition
that $\Sigma_1$ and $\Sigma_2$ are disconnected.

 It is worth spelling out condition (2)
as its consequences are far-reaching. This condition is that the
group of diffeomorphisms of a $(d-1)$-manifold $\Sigma$ is represented
by linear transformations in $V(\Sigma)$, and that any diffeomorphism of
$\Sigma=\partial M$ {\it which extends over} $M$ leaves the partition function
$Z(M)$ invariant. The same conditions hold for diffeomorphisms between
different manifolds.

Since any isotopy of $\partial M$, also called a small diffeomorphism,
extends over any manifold $M$,
an immediate consequence of
this is that all partition functions are invariant under the action of
isotopies. Translated into the language of quantum cosmology, the
wavefunction of the universe satisfies the equations known as the
momentum constraints.

An isotopy of $X$ is a diffeomorphism
$$X\times[0,1]\to X\times[0,1]$$
which preserves the `time' parameter (so that the image of $(x,t)$ lies in
$X\times\{t\}$) and leaves fixed all the points on one end, $(x,0)\mapsto
(x,0)$. The mapping of the other end, $X\times\{1\}$, is the `small'
diffeomorphism of $X$.
The reason that an isotopy of $\partial M$ extends over $M$ is that there is
always a `collar' neighbourhood of $\partial M$ diffeomorphic to the cylinder
$\partial M\times [0,1]$. The isotopy extends over the collar by definition,
and
extends to the other points of $M$ by the identity mapping.

If the boundary of a $d$-manifold $C$ is regarded as separated into two
disjoint closed parts,
$$\partial C=\Sigma_1\cup\Sigma_2,$$
then the vector $Z(C)\in V(\Sigma_1)\otimes V(\Sigma_2)$ can be regarded
as a matrix of transition amplitudes
$$V(\overline{\Sigma}_1)\to V(\Sigma_2).$$
The normal rules of quantum mechanics, (1), require that these multiply as
matrices when the corresponding manifolds are glued together.
A technicality to mention is that for this to work, the $(d-1)$-manifold
$\overline{\Sigma}_1$ is $\Sigma_1$ with the opposite orientation, and its
state space is the dual space to that of $\Sigma_1$. The bra and ket notation
can be used to keep track of these.

Now, gluing an extra collar $\partial M\times [0,1]$ onto $M$ does not
change the topology.
Therefore one has the linear equation
$$Z\bigl(\partial M\times [0,1]\bigr)Z(M)=Z(M)$$
which shows that the vector $Z(M)$ is a unit eigenvector of the matrix
$Z(\partial M\times [0,1])$. In the quantum cosmology, this equation
is the statement that the wavefunction satisfies the Hamiltonian
constraint. The transverse vector field which normally accompanies the
Hamiltonian constraint is simply an infinitesimal version of the diffeomorphism
$$M\cup \bigl(\partial M\times [0,1]\bigr)\to M.$$

The Hamiltonian constraint is normally written
$$H Z(M)=0,$$
the operator $H$ corresponding to an infinitesimal transverse deformation.
The usual relationship between the Hamiltonian and the evolution operator
is exponentiation, so one has the equation \refer{Atiyah book}
\refname\Atiyahbook
$$e^{-H}Z(M)=Z(M).$$
{}From the analogy with canonical quantisation,
one might expect one equation for every infinitesimal transverse deformation.
They are however all the same, since any two transverse vector fields are
related by an isotopy.

A further comparison with quantum cosmology shows some important points of
physics. Although the no-boundary proposal is formulated for a particular
manifold, $B^4$, quantum
cosmology allows the consideration of the functional integral on
any manifold. One has therefore in mind a very specific construction in which
$V(\Sigma)$ is the vector space of functions on the space of Riemannian
metrics on $\Sigma$. Therefore one has right away a link with geometry and
the necessary information to construct observables with a prescribed
geometric meaning. This information is not present in the general framework
of TQFT, but must be sought in the further structure which is provided by
particular examples.

However it is remarkable that one can make a lot of progress using the
general properties of TQFT alone, constructing particular wavefunctions
and showing that they satisfy the momentum and Hamiltonian constraint
equations. One might ask how this has been done, given that constructing
solutions to these equations in the framework of canonical quantisation
has been considered an important and apparently intractable problem. The
answer is that the difficulty has been transformed into the combinatorial
problem of constructing invariants. This problem is also difficult; but it
is at least within the domain, and technology, of modern mathematics.

The primary problem is in fact the construction of invariants for closed
manifolds, and the state spaces and partition functions for manifolds with
boundary are usually obtained as a by-product.

Gluing two cylinders $\Sigma\times [0,1]$ on top of each other, one
obtains $\Sigma\times [0,2]$, which is of course diffeomorphic to
$\Sigma\times [0,1]$. This translates into a matrix equation
$$Z\bigl(\Sigma\times [0,1]\bigr) Z\bigl(\Sigma\times [0,1]\bigr)=
Z\bigl(\Sigma\times [0,1]\bigr).$$
In other words the cylinder partition function is a projection operator on
the state space of $\Sigma$. It will be useful to use a shorthand notation
for this projector,
$$P_\Sigma\equiv Z\bigl(\Sigma\times [0,1]\bigr).$$

Apart from showing that the eigenvalues of $H$ are $0$ and $\infty$ only,
this raises the possibility of replacing each $V(\Sigma)$ in the theory
with the image of the corresponding cylinder projector. From the point of
view of the axiomatic framework of TQFT, which is all that has been
introduced so far, one may as well do this. Since all partition functions
lie in these images, the theory has the same properties and the same
invariants.
However, particular examples of TQFTs have further structure, for which
the fact that the $V(\Sigma)$ is larger than the image of the cylinder
projector is important.

An observable is represented by an operator in a state space, $V(\Sigma)$.
The observable $\Cal O$ can be called a global observable if the operator acts
non-trivially only in $\image P_\Sigma$, so that
$$\Cal O P_\Sigma=P_\Sigma\Cal O=\Cal O.$$
Otherwise it is called a local observable.

The argument given above that partition functions are invariant under the
action of isotopies can be applied to the manifold $\Sigma\times[0,1]$
itself to show that all elements of $\image P_\Sigma$ are invariant under
isotopies. This means that global observables cannot refer to a particular
subset of $\Sigma$ and they are indeed only sensitive to the global topology
of $\Sigma$. By contrast, in particle physics most observables can
be localised to one small region. Therefore the global observables are very
limited and do not capture the full content of dynamics.

Global observables can be constructed using only the machinery introduced
so far, as the examples described in the next section demonstrate, and
allow for a relatively simple and self-contained discussion of quantum
theory.
Local observables will be discussed later.

\head IV. Global Observables \endhead
The formalism which has been introduced contains some of the mathematical
elements of quantum theory. A quantum theory proper must have a method
of calculating probabilities, together with a statement of the physical
objects or processes that these correspond to. To meet the first point,
some projection operators will be defined, which enables a discussion
of the relations between the probabilities which are determined by them.
These operators are loosely referred to as observables, although the term
more strictly refers to the physical object or process to which these are
said to correspond.

If $C$ is a manifold with boundary
$\Sigma\cup \overline{\Sigma}$, then $Z(C)$ is an operator
$$V(\Sigma)\to V(\Sigma)$$
just as for the case $C=\Sigma\times[0,1]$ considered previously.
The simplest case to consider is when $Z(C)$ is a multiple of a projector, so
let $C$ be this special type of manifold.

Suppose that, as in quantum cosmology, $M$ is a space-time manifold with
boundary, and let $\partial M$ be identified, for the moment, with $\Sigma$.
The observable labelled by $C$ acts on partition functions $Z(M)$ in
$V(\Sigma)$
by gluing $C$ on to the boundary of $M$. It is a global observable because
$$C\cup_\Sigma\bigl(\Sigma\times[0,1]\bigr)\cong C,$$
in which the cylinder can be glued to $C$ at either end.

This situation can be generalised. The closed $(d-1)$-manifold $\Sigma$
need only be one component of the boundary of $M$. More generally,
it could be any embedded submanifold of $M$, called a hypersurface, such that
cutting $M$ along $\Sigma$
produces a manifold $M_\Sigma$ with two extra boundary components, $\Sigma$ and
$\overline\Sigma$. The state space for $M_\Sigma$ is
$$V(\Sigma)\otimes V(\overline\Sigma)\otimes V(\partial
M)\cong\End(V(\Sigma))\otimes V(\partial M),$$
$\End(V)$ being the set of linear operators on $V$.
The observable labelled by $C$ also acts in the space $V(\Sigma)$. Gluing $C$
onto $M_\Sigma$ on both boundary components $\Sigma$ and
$\overline\Sigma$ produces a manifold
$$M_\Sigma C=M_\Sigma\cup_{\Sigma\cup\overline\Sigma}C$$
which is $M$ with $C$ sewn into it. The gluing rules give
$$Z(M_\Sigma C)=\tr_{V(\Sigma)}Z(M_\Sigma)Z(C)\in V(\partial M),$$
taking the trace over $V(\Sigma)$. In the language of physics,
$C$ operates on a subsystem which is coupled to other systems.


One of the shifts in thinking about quantum theory which is involved is that
some of the considerations are only valid for a particular quantum state.
Here, the observable labelled by $C$ makes a difference by replacing the
state vector $Z(M)$ with $Z(M_\Sigma C)$. However this is not defined as the
result of applying a linear operation to the space $V(\partial M)$. If $M'$
is another manifold with the same boundary, $\partial M$, then the action
of $C$ on $Z(M')$ is not defined because one does not know, in a general
situation,
how $\Sigma$ should be inserted in $M'$. In the special situation where
$\Sigma$ lies in $M$ as a level hypersurface of a collar neighbourhood of the
boundary $\partial M$ (i.e., parallel to the boundary), then the usual rule is
recovered,
$$ Z(M_\Sigma C)=Z(C)Z(M),$$
with $Z(C)$ acting as a linear map on the vector $Z(M)$. This is essentially
the
same as when $\Sigma=\partial M$.

How many of these subsystems are there? If $\Sigma'$ is a second embedded
hypersurface in $M$, then there is an equivalent observable on $\Sigma'$
if there is a diffeomorphism
$$f\colon M\to M,\quad\quad f_{|\partial M}=\identity$$
which takes $\Sigma$ to $\Sigma'$. Then inserting $C$ at $\Sigma$ is
equivalent to inserting $C$ at $\Sigma'$, provided $\partial C$ is identified
with $\Sigma'\cup\overline{\Sigma'}$ using $f$.

The question of the number of inequivalent embeddings of a given
$(d-1)$-manifold in $M$ gets rapidly complicated as the dimension
increases through $d=2,3,4$. In dimension two, there is only one way to
embed a circle in $S^2$, and a finite number of ways of embedding it in a
closed surface, enumerated by the topologies of the pieces
obtained by removing the circle. In dimension three, there is still only
one way to embed $S^2$ in $S^3$, but an infinite number of ways to embed
a torus - taking a tube around any knot, for example. However, any surface
embeds in $S^3$ in a unique way if the complement is required to be
two handlebodies. In dimension four, it is not even known if the
embedding of $S^3$ in $S^4$ is unique. This is known as the Sch\"onflies
conjecture.

In the usual quantum mechanics, operators are conveniently regarded as
being defined at one particular instant of time, with a sequence of
interactions being represented by the {\it multiplication} of the
corresponding operators in the order given by the causal structure, or time
coordinate. Therefore, the essential topological feature of the usual quantum
mechanics is the time ordering of observables. This has been regarded as
a difficult feature of quantum gravity, since the causal structure of a
space-time is a feature of the classical metrics, which arise as the spectra of
quantum observables, and is therefore not available for the overall structure
of the quantum theory. Another puzzling feature, on physical grounds,
is the postulate that the observables should form an algebra. It may not
make sense to consider two interactions which happen in one time order as
happening in the opposite time order, yet in an algebra the product
is always defined in both orders.

In a TQFT, this information is specified by choosing a particular hypersurface
$\Sigma\subset M$, together with some further information concerning
the observable in $\Sigma$. This is identifying some topological objects
(here $\partial C$) on $\Sigma$. Thus specifying a {\it place} in the
manifold replaces specifying a time ordering in the usual theory.

If $M=\Sigma\times[0,1]$, and the hypersurfaces considered are just
$$\Sigma\times\{ t\} \subset \Sigma\times[0,1]$$
for different values of $t$,
then the normal time ordering is apparent in the TQFT.
The gluing rules of the TQFT imply that, in this situation, the observables
are indeed multiplied as operators in the corresponding order. The
multiplication of an ordered set of operators is merely a particular
case of the dynamics of a topological theory. These conclusions
are also justified for a general manifold $M$ if all observables considered
are in a collar neighbourhood of the boundary.

However in the general situation, the linear time ordering for the observables
may not exist, and in general it may not make sense to swop the places at
which two different observables are defined. For this more general notion
of quantum theory, it is necessary to discuss the probabilities that the
theory determines directly. An example of a theory in which this more
general discussion makes sense will follow.

\subhead\nofrills A. Hermitian theories \endsubhead

 For this discussion one wants to have Hermitian inner products,
for the same reason
as in quantum mechanics, that the square of a vector should be a probability.
A Hermitian theory is one in which each state space has a Hermitian inner
product, which, considered as an (antilinear) map
$$V(\Sigma)\to V(\overline{\Sigma})$$
takes any partition function $Z(M)$ to $Z(\overline{M})$, for $\partial M=
\Sigma$.
In the bra-ket notation, $Z(M)$ is written $\C M\ket$ and $Z(\overline
M)$ is written $\bra M\C$. In the Hermitian theory, $\bra N\C M\ket$
is the Hermitian inner product of $Z(N)$ with $Z(M)$. Therefore for the
double of any manifold, obtained by gluing two copies across the boundary,
$$Z(M\cup_{\partial M}\overline M)=\bra M\C M\ket \ge0.$$
Also, if $P$ is a closed manifold, then $Z(\overline P)
\in\Bbb C$ is the complex conjugate of $Z(P)$. This is a special case
of the more general fact that for the manifold $C$, the matrix
$Z(\overline C)$ is equal to the Hermitian adjoint $Z^+(C)$ (To show this,
one has to use the natural assignments of Hermitian inner products to the dual
of a state space and the tensor product of two state spaces.).

\subhead\nofrills B. Probabilities \endsubhead

A discussion of probabilities is essential to give the theory some meaning
as a theory of {\it physics}\/. It is worth starting with a discussion
of standard quantum mechanics in a single state space. Let $\pi_1$ be an
orthogonal projector in the inner product space $V(\partial M)$.
Then the usual discussion of quantum mechanics is to associate $\pi_1$ to a
proposition $X$. If one wishes to assert $X$ as a fact, then a state vector,
say
$\C M\ket$, is simply replaced by the vector $\pi_1\C M\ket$ for the purposes
of future discourse.

If, however, one wants to discuss the relative merits
of $X$ and \NOT $X$, which is assigned the projector
$1-\pi_1$, then the probabilities
$$p_1={\bra M \C \pi_1\C M\ket\over \bra M\C M\ket} \hskip 30pt\text{and}\hskip
30pt
 p_{\bar 1}={\bra M \C 1-\pi_1\C M\ket\over \bra M\C M\ket}$$
are used.
It is obviously essential that these numbers are not negative, and add up to 1.
Assuming the
denominator is not zero, this is because the numerators are both the square of
a vector, which implies $p_1,p_{\bar 1}\ge0$, and because of the condition
$$ \bra M\C M\ket=\bra M \C \pi\C M\ket+\bra M \C 1-\pi\C M\ket.$$
which implies
$$ 1=p_1+ p_{\bar 1}.$$
This is the most elementary example of a consistency condition for the
probabilities, which is satisfied automatically, as $\pi_1$ is a projector.

Now suppose $\pi_2$ is a second projector, which need not commute with
$\pi_1$, and define $p_2$, $p_{\bar 2}$ in the same way using the state $\C M
\ket$. According to Griffiths \refer{}\refname\Griffiths, the numbers
$$\aligned p_{12}&={\bra M \C \pi_1\pi_2\pi_1\C M\ket \over \bra M\C M\ket}\\
\\
p_{\bar12}&={\bra M \C (1-\pi_1)\pi_2(1-\pi_1)\C M\ket\over \bra M\C
M\ket}\endaligned
\qquad\qquad
\aligned p_{1\bar2}&={\bra M \C \pi_1(1-\pi_2)\pi_1\C M\ket
\over \bra M\C M\ket}\\ \\
p_{\bar1\bar2}&={\bra M \C (1-\pi_2)(1-\pi_1)(1-\pi_2)\C M\ket
\over \bra M\C M\ket}\endaligned$$
determine a joint probability distribution on the four-point space
$\{12,1\bar2,\bar12,\bar1\bar2\}=\{1,\bar1\}\times\{2,\bar2\}$ if a
consistency condition is satisfied. This is the condition
$$p_2=p_{12}+p_{\bar12}.$$
The three other analogous relations for $p_{\bar 2}$, $p_1$, $p_{\bar 1}$
follow automatically. These conditions are the conditions for the
compatibility of the
marginal distributions for $\{1,\bar1\}$ and $\{2,\bar 2\}$ determined by
$p_1$ and $p_2$ with this joint probability distribution. As explained by
Omn\`es \refer{}\refname\Omnes, this condition, and its generalisation to more
than two
propositions, is a necessary and sufficient prerequisite for consistent
reasoning
in a quantum theory.

The interpretation of $p_{12}$ in quantum
theory is that it represents the probability of the proposition $\pi_1$
followed
by $\pi_2$ at a later time. The joint probability distribution is used to
discuss the correct inferences which can be drawn connecting different
propositions at different times. These are the implications which hold
with probability one. Quantum theory is a language whose elements
refer to the many different propositions which may be considered and whose
content is the many relations between them. This is the sense in which the
consistency conditions are essential; the individual probability distributions
have a very limited scope of application on their own.

Omn\`es argues that all true statements about quantum systems fall
into the class of correct implications which can be made using this probability
calculus.
This argument is not provable in a mathematical sense, but has an experimental
status, like Church's thesis on computability:
all situations which have been analysed have indeed followed this pattern.
It is fairly easy to demonstrate, on the other hand, that the failure of the
consistency conditions signals the inapplicability of Boolean logic and the
possibility of logical paradoxes. It is also the essential feature of
quantum interference phenomena.

An important point to note is that the consistency condition depends on the
quantum state. Indeed, requiring it to hold for all states in the state space
would force the projectors $\pi_1$ and $\pi_2$ to commute, which, if required
universally, would rule out all quantum phenomena.
 The fact that Boolean logic applies for some quantum states in some
situations where the observables do not commute is a relatively recent
discovery. It invalidates the earlier assumption that a modification of
Boolean logic is necessary to discuss any quantum phenomena at all.

This analysis extends to probabilities which can be calculated in a Hermitian
topological quantum field theory. As explained above, the operator $Z(C)$
can be considered to act in the vector space $V(\Sigma)$ if the boundary of
$C$ is identified appropriately with $\Sigma$. Equally it makes sense to
act with a linear combination of such operators, even if the geometrical
interpretation is less clear. Thus if $Z(C)$ is not a projector, some
polynomials in $Z(C)$ may be. Let us continue with the case mentioned above,
where $Z(C)$ is a multiple of a projector, which will hold in the specific
example to be introduced below. The projector is
$$\pi=\lambda Z(C), \qquad \lambda\ne0\in\Bbb C,$$
acting in the space $V(\Sigma)$, for the hypersurface $\Sigma\subset M$,
which need no longer be the boundary of $M$ or parallel to it.
The complementary observable is
$$ 1-\pi= 1 - \lambda Z(C).$$
Accordingly, the two vectors obtained from the action of these two operators
are
$$\C \pi M\ket=\lambda\C M_\Sigma C\ket  \qquad\text{ and }
\qquad
\C (1-\pi) M\ket=\C M\ket - \lambda\C M_\Sigma C\ket.$$
The notation $\C \pi M\ket$ has been chosen to emphasise that $\pi$ does not
act in $V(\partial M)$; it is only in the special case where
$\Sigma$ is parallel to $\partial M$ that one can write
$$\C \pi M\ket=\pi\C  M\ket.$$
The probabilities are defined as the normalised square of these vectors
$$p={\bra \pi M \C\pi M\ket\over \bra M\C M\ket} \hskip 30pt\text{and}\hskip
30pt
 \bar p={\bra (1-\pi) M \C (1-\pi) M\ket\over \bra M\C M\ket}.$$
There is a consistency condition for these probabilities, namely that
$p+\bar p=1$. This can be written in the form
$$\bra \pi M \C\pi M\ket=
{1\over 2}\Bigl(\bra \pi M \C M\ket+\bra  M \C\pi M\ket \Bigr),$$
or as
$$ \operatorname {Re}\bra(1-\pi)M\C \pi M\ket=0.$$
The later form has a state space interpretation which is that the
two vectors are orthogonal in the real-linear form given by the
real part of the Hermitian inner product.
The condition is analogous to the Griffiths consistency condition for
ordinary quantum mechanics, with the non-trivial topology of the manifold
playing the role of the interposition of the projector $\pi_2$.

This formalism can be extended to consider the joint probability distributions
determined by a number of projectors each acting in a different place in
the manifold. Several consistency conditions will arise. These conditions are
that the joint probability distributions are compatible with each of the
marginal distributions, which are calculated
in the same manner from the quantum theory.

The formalism has the property that if a number of observables
act in parallel hypersurfaces in a submanifold isomorphic to
$\Sigma\times[0,1]$, then the calculation for global observables will reduce to
the usual quantum
mechanical multiplication of projectors. In particular, it is possible to show
that repeating the same proposition is reliable. Let $X_t$ be
a proposition which is represented by the projector $\pi$ acting in the
state space associated to $\Sigma\times\{t\}$. Then
$$p(X_t\cap X_s)=p(X_t)=p(X_s),$$
due to the fact that $\pi P_\Sigma \pi=P_\Sigma \pi=\pi P_\Sigma$.
This means that
$$p(X_t|X_s)={p(X_t \cap X_s)\over p(X_s)}=1,$$
and according to Omn\`es interpretive rule for quantum mechanics, the logical
implication
$$X_s \Leftrightarrow X_t$$
therefore holds.

\subhead\nofrills C. Semantics \endsubhead

The previous discussion can be completed by describing a framework for a
semantics. This will allow some general conclusions to be drawn about the scope
of the theory.

What remains is to discuss the correspondence between the physical objects or
processes, and the mathematical entities consisting
of certain specific projection operators acting in certain state spaces.
This is a difficult point in all discussions of quantum gravity.

In more conventional
physics, the semantics is constructed from the specification that there are
some physical objects which behave in the manner of classical deterministic
dynamics. The classical example is `the particle is at a position $x\in[a,b]$'.
Let us call this proposition $X$.
{}From this statement one can construct, in particle physics theories,
a projector in the particular Hilbert space of interest. These projectors,
for different subsets of $\Bbb R$, together with the analogues for momentum
observables, form a privileged subset of the set of
all projectors in the Hilbert space.

 Such a statement is made in a particular context, which the usual theory
presupposes.
This is that there are also a number of macroscopic reference bodies, whose
existence delineates the classical `space', $\Bbb R$, and that the particle is
in fact correlated with these macroscopic bodies. This context is a matter
of physical fact, rather than mathematical formalism. For example, a particle
emitted by radioactive decay of a nuclear state has a position variable
which is in fact correlated with the atom which has emitted it. The atom is,
in turn part of a lump of solid matter, which is fixed in a laboratory. This
laboratory contains macroscopic measuring devices which serve to define the
interval $[a,b]$. Finally, the description of the particle
by its wavefunction can be tested by a measurement interaction which alters
the positions of the macroscopic bodies.

The existence of the macroscopic bodies, and the initial conditions, is
asserted as fact, say by a collection
of statements $Y$. This has a different status to the proposition $X$, in that
$Y$ is asserted as fact, and the possibility of \NOT $Y$ is not entertained, or
even defined. On other hand, $X$ and \NOT $X$ are both defined and ascribed
probabilities which satisfy a consistency condition when they are used. This
consistency condition ensures that $X$ and \NOT $X$ relate to each other in the
way that is usually meant.
 The future discourse may introduce further propositions of this type, and the
truth of $X$, \NOT $X$, or combinations of these with other propositions
is a matter of calculation and logical inference. The facts $Y$ on the
other hand are simply a prerequisite for the debate and their truth is
not open to discussion.

The theory of a single particle is normally written without the need to
include macroscopic reference bodies explicitly in the quantum mechanical
formalism. The correct procedure, as ever, is to include only the
least amount of information about these other bodies in the description of the
particle. This information is precisely the commonly agreed meaning
of the privileged sets of projectors. Other projectors in Hilbert space, such
as
those specifying momenta, can be introduced, but only in as far as they
have a particular specified relation to the reference bodies. In this way, the
rather featureless
geometry of Hilbert space is replaced by the geometry of the phase space of
a particle.

Quantum theory does supply an interpretation for the situation where sets
of projectors and a quantum state fail to satisfy the Griffiths consistency
condition. For example, in the double slit experiment, the proposition $X$
could be that the particle has gone through one of the slits. As is well known,
the probability sum rules in this experiment are not satisfied, so that
the probabilites $p(X)$ and $p(\text{\NOT }X)$ calculated from quantum theory
do not sum to $1$ when
normalised with the probability calculated for the double slit experiment
where the particle is free to travel through either slit. This logical
paradox is resolved by the necessity to insert a measuring device in front
of one of the slits to test the truth of $X$ or \NOT $X$. Therefore, the only
situations where $X$ or \NOT $X$ have any logical consequences are where the
physical facts $Y'$ actually differ from the
asserted facts $Y$ for the double slit experiment without the determination
of $X$ or \NOT $X$, according to the presence or absence of the measuring
probe. In other words, the theory asserts that a discrepancy in
the consistency condition is due to the difference in the asserted facts $Y$ or
$Y'$, and this is due to an actual physical cause. In established quantum
physics, this cause can be described within the quantum formalism, and
consistency is restored.

There is an analogy in Newtonian mechanics. Suppose that the known forces on
a body do not account for its acceleration. Then according to Newton's laws,
there must be another force on the body, which is due to the action of another
physical object which has not yet been included in the formalism. Including
all bodies which act does eventually resolve the discrepancy.

Finally, conventional quantum mechanics allows one to assert any prior facts
$Y$ which are not absolutely contradictory. This means that the quantum state
vector is not precisely zero. In that case, the state vector can
be normalised by dividing by its norm, which is the same as normalising the
probability measure to give a total measure one. This means one can add any
fact to $Y$ asserted by the action of a projector, as long as the resulting
state vector is still non-zero. This wide choice of quantum state appears to
rule out any explanation for any objective facts, common to all observers,
within conventional quantum theory.

The semantics for quantum gravity is necessarily more abstract. In the
preceding discussion of global observables, the various choices of manifold
$M$,
hypersurfaces $\Sigma$ and observables, labelled by $C$, need to be assigned a
physical counterpart. By deliberate construction, all three of these are
topological objects. Therefore, I would like to propose a general
principle that the topology of the {\it physical} situation is reflected
in the topology which enters into the {\it mathematical} description of the
state vector and the observables. This means that the projectors which occur in
the theory have a very particular construction in terms of the topology of
manifolds. The topological construction singles out a priveleged set of
projectors, which plays the role that the set of position and momentum
projectors enjoys in particle physics.

The way in which this principle is applied will depend on the particular model,
and the ideas at present are rather vague. The global observables are rather
limited, and a more thorough investigation would start with the
properties of local observables. However, some general remarks can be made.

Following the ideas of quantum cosmology, one could propose that the manifold
$M$ represents both space-time and the initial conditions. By analogy with the
quantum mechanics of a particle, the choice of $M$, and
thus the initial state vector, $\C M\ket$, is one of the asserted facts $Y$,
and one is free to choose any $M$ so long as $\C M\ket \ne 0$. As the global
observables change the topology of $M$, this confirms that they are truly
global in the space-time
sense, as opposed to having a localised effect in a small region.

The introduction of propositions whose projectors satisfy consistency
conditions
has immediate implications for the boundary of $M$. As noted above, one form
of the condition for one observable is that the two vectors in $V(\partial M)$
are orthogonal. This feature persists in the consistency conditions for several
observables, with the number of mutually orthogonal vectors increasing with the
number of consistent observables. This means that unless the vectors are zero,
the dimension of $V(\partial M)$ has to be sufficiently large to support
 the observables. In particular a closed manifold $M$ has little use. It also
implies that the observables which are required must have a certain
dynamical connection with the boundary $\partial M$.

One interpretation of these conditions is that $\partial M$ represents the
`present instant of time' and that the vectors in $V(\partial M)$ which are
determined by various choices of observables represent a memory of past facts.
At least if the vectors are orthogonal in the Hermitian inner product, a
stronger condition, then they are in principle distinguishable in a
deterministic way by operations in $V(\partial M)$. Such a dynamical connection
with the boundary is also required for a theory of {\it measurements}, where
facts are correlated with memory devices.

These general remarks on the framework for a physical interpretation of the
theory show that it is reasonable to construct the semantics for quantum
gravity in a fashion which is analogous to, but more abstract than, the
standard quantum mechanics for particles. The idea is to replace the classical
world (or in more sophisticated treatments, the approximately classical world)
of rigid bodies in space-time with some prior notions of topology. In a sense,
one can say that quantum gravity is, amongst other things, a `quantum theory
without large objects'. This means that the description of a small object
should not ultimately rest on the existence of large objects such as rigid
bodies to construct the semantics for the requisite observables.

 Some more ambitious goals are sometimes mooted for quantum gravity, often
under the name of quantum cosmology. One such project is to use the extra
information
in the initial state vector, namely its norm
$$\bra M \C M\ket,$$
as a prior probability for the `occurence' of $M$. The idea is to take a `one
world' view in which every possible fact or potentiality does occur in the one
universe with some frequency or probability. The problem with this is that the
event space (in the sense of probability theory) is not clear, with
propositions such as `space-time is not $M$' having no clear translation into
the quantum formalism. A second problem with this project is that the quantum
formalism appears to treat the observer, namely the
person or object which is entertaining the propositions, as outside of the
quantum system described by the TQFT. This type of observer dependency may be
described by a framework in which quantum states of the usual kind are relative
to observers, but some type of abstract absolute quantum theory is retained as
an overall description \cite{\reference{Everett}\refname{\Everett},
\reference{Rovelli}\refname\Rovelli}. However such a framework has not been
developed in the context of topological quantum field theory.

An even more ambitious project is to assert the existence of common or
objective facts. This requires not only the selection of some universal initial
state but also some restriction on the allowed propositions, with say, the
consistency condition allowing only  a single set of propositions \refer{Dowker
and Kent}\refname\Dowker. At the present stage it is too early to say whether
these doctrines have a precise formulation within the framework which is
discussed here.

\subhead\nofrills D. An example\endsubhead

A simple example can be presented in dimension three (corresponding to
two-dimensional `space', 3=2+1). This example is for a theory which satisfies
a condition on connected sums. The connected sum of two connected manifolds,
$M$ and $N$,
is obtained by removing a ball $B^3$ from each, and joining $M$ and $N$ across
the extra boundary components $S^2$ so created on each one.
 The balls can be taken to be small neighbourhoods of some arbitrarily chosen
interior point on
each. The connected sum is denoted $M\# N$.

The condition that the theory satisfies for the example to work is
$$Z(S^3) Z(M\# N)=Z(M)\otimes Z(N).$$
The scalar $Z(S^3)$ multiplying $Z(M\# N)$ can be thought of as correcting for
the two balls which are removed from $M$ and $N$.
This condition is satisfied for the Turaev-Viro model discussed in the next
section, for example.

The observables are obtained by the operation of surgery. Suppose that $b$ is
a band $S^1\times [0,1]$ embedded in the 3-manifold $M$. The band can be
thought
of as lying in a small neighbourhood of the circle $S^1\times\{0\}$.
This circle has a neighbourhood in $M$ which is a small solid tube with
topology
$S^1\times B^2$. The band gives a preferred way of identifying the boundary
of this tube with a standard torus $S^1\times S^1$, at least up to isotopy,
which is what counts. Surgery is the operation of removing the tube and
re-inserting it after swopping the variables
$$(x,y)\mapsto (y,x)$$
on its boundary, $S^1\times S^1$. The new manifold obtained from $M$ by surgery
on $b$ is denoted
$$\surgery(M,b).$$
The other end of the band, $S^1\times\{1\}$, can be taken to lie on the
boundary of the tube.
After surgery, this curve is contractible, whereas before the surgery it may
not be.

The observable is constructed from a surface (2-manifold) $\Sigma$ and a band
$b\subset\Sigma$. The band $b$ can also be embedded in the
middle of the cylinder $\Sigma\times[0,1]$, say
$\bbar=b\times\{1/2\}\subset\Sigma\times[0,1]$.
Now let
$$C=\surgery(\Sigma\times[0,1],\bbar).$$
With the above condition on connected sums, one can show that $Z(C)$ is a
multiple of a projector $\pi$, in fact,
$$Z(C)^2={ Z(S^2\times S^1) \over Z(S^3)} Z(C).$$
This is because if $C$ is glued to a copy of itself by one end, then the band
$\bbar_1$ in one copy can be slid up the cylinder until it lies parallel to the
other band $\bbar_2$. Then after the surgery on $\bbar_2$, $\bbar_1$
can be slid to a
band which lies in a small ball neighbourhood of one point, an `unknot'.
Finally, surgery on the band $\bbar_1$, now in the small ball, gives the
connected sum of $C$ with $S^2\times S^1$.

Following the above formalism for observables, suppose that $M$ is a
3-manifold and $\Sigma\subset M$. Then $C$ can be inserted in $M$ at $\Sigma$
to give $M_\Sigma C$. However there is an isomorphism
$$M_\Sigma C\cong\surgery(M,b),$$
and the later does not depend on the choice of $\Sigma$. Therefore the
consistency condition is satisfied if a very simple topological condition
is satisfied. If $b$ can be deformed to lie in the boundary of $M$, then one
has
$$\bra \pi M\C\pi M\ket=\bra  M\C\pi M\ket,$$
sometimes called the strong consistency condition. This follows from the same
argument that shows that $Z(C)$ is a multiple of a projector.

There are an infinite number of inequivalent such observables, but typically
only a few of them will satisfy this consistency condition. Picking a surface
$\Sigma\subset M$, one can see that there will be some bands in $\Sigma$
which will satisfy the condition, but others which may not, unless of course
$\Sigma$ is parallel to the boundary of $M$, when all will be consistent.
This situation is analogous to the situation in quantum mechanics, where
for intermediate times in a Griffiths history, the typical observable will not
be consistent with the rest of the history, but some special choices will be.

This verifies the existence of quantum interference phenomena in this example.

 \head V. Examples of TQFTs in dimension three \endhead

The main examples of TQFTs were preceded by some theories which have a similar
construction but turn out to be defined only for some manifolds. This is for
the usual reason in quantum field theory, that some infinite sums fail to
converge. I shall describe these theories first, before turning to the
mathematically watertight TQFTs, mostly based on quantum groups.

\subhead\nofrills A. Three-dimensional gravity \endsubhead

A state-sum model for a partition function is familiar from statistical
mechanics, where it is defined by summing the Boltzmann weights, given by the
exponentials of an energy, over a set of states or configurations for a system.
The key features
are that the states are determined by variables which are distributed
over a lattice, and the weights are the product of factors defined locally
on each lattice site.

The main ingredients for a state-sum model defining quantum gravity in
dimension three were assembled by Ponzano and
Regge\refer{}\refname\ReggePonzano.
The partition function is defined by a sum which can be thought of as a
discrete version of a path integral. The variables are on the edges of a
triangulation of the 3-manifold, somewhat similar to a lattice gauge theory.
However, the theory differs from the lattice gauge theory in that a change of
triangulation (for example a refinement) in the interior of the manifold
is supposed to give {\it exactly the same} partition function. This is an
expression of the topological invariance.

The variables are summed over an infinite
set of values, and in general this sum will diverge. Ponzano and Regge
suggested a regularisation scheme which clearly gives a finite answer for the
simplest cases, but the properties of this regularisation have not been
developed in general. There are some triangulations of some 3-manifolds
for which the regularisation is not needed, namely those for which the
boundary is `sufficiently large' that the fixed boundary data place constraints
on the interior variables appearing in the state sum, which constrain these
to a finite range of values. Therefore, without needing to consider the
properties of a regularisation, one can say that the theory is partly defined.

The variable associated to each edge ranges over the set of representations
of the Lie group $\SU(2)$. This set can be identified with the non-negative
integers. In the analogy with the path integral, these integers are regarded
as lengths for these edges, thus defining a metric tensor on the manifold,
which is flat on each simplex. This affords the most direct link between
topological theories in three dimensions, and quantum gravity. As remarked
earlier,
it depends on some extra properties which this model has, in addition to
the general properties investigated in the previous sections.

There is a Boltzmann weight for each tetrahedron, given by the
value of the $6j$-symbol at the six representations of $\SU(2)$ on the edges
of the tetrahedron. Ponzano and Regge gave arguments for a remarkable formula
for an asymptotic limit for this Boltzmann weight, as the integers on
the edges approach infinity.
The limit is that the $6j$-symbol tends to the exponential of the Einstein
action for the metric \cite{\ReggePonzano, \reference{Schulten and
Gordon}\refname{\Schulten}, \reference{Biedenharn and
Louck}\refname\Biedenharn, \reference{Barrett and Foxon}\refname\Foxon}, as
given by Regge calculus \refer{Regge}\refname\Regge. There considerations were
all heuristic, and it is still an open problem
to provide a precise statement of the asymptotic limit, and proofs.

Witten provided a definition of quantum gravity in three dimensions starting
from the Einstein action and the path integral \refer{Witten topology
changing}\refname \WittenTop. The evaluation of the path integral boils down to
a formula involving the
Ray-Singer analytic torsion, a formula which is only finite in some
circumstances. Some explicit calculations have been done by Carlip and
Cosgrove\refer{}\refname \Carlip. The first circumstantial piece of evidence
that this theory
is the same as Ponzano-Regge is that the manifolds for which this is
well-defined, such as handlebodies, are precisely those for which the
Regge-Ponzano partition function can be calculated via a finite sum.
The fact that both theories have a construction (approximately or exactly)
from the Einstein action is the second piece of circumstantial evidence. The
third piece
of evidence is that both theories can be regarded, heuristically, as a limit
of a well-defined theory, described below, for which there is a theorem
asserting that the two theories are equivalent.

An argument which would settle the issue once and for all would be a
regularisation of the Regge-Ponzano theory which gave the combinatorial
definition in terms of the Reidemeister torsion. The equivalence would then
rest
on the equality of Reidemeister and Ray-Singer torsion. However such an
argument has yet to be carried through.

\subhead\nofrills B. Topological quantum field theories \endsubhead

The Ponzano-Regge theory
requires very little modification to determine a bona fide TQFT. The Lie group
$\SU(2)$ is replaced by the quantised enveloping algebra of the Lie algebra
$\Sl(2)$ at a root
of unity. This manoeuvre ensures that there is a {\it finite} set of
representations for which a completeness relation still holds. At a stroke,
the infinite sums which plague three-dimensional gravity are replaced by finite
sums and the theory satisfies all of Atiyah's axioms for a TQFT. This is
the theory of Turaev and Viro\refer{}\refname \TV. The theory was subsequently
generalised to other quantised enveloping Lie algebras (quantum groups)
\cite{\reference{Barrett and Westbury}\refname\Westbury, \reference{Turaev
book}\refname\TuraevBook}. Although the formalism generalises easily enough,
the non-trivial part of this work is to check that these other quantum groups
do satisfy all of the relevant conditions. There are also models which
apparently
have nothing to do with quantum groups \refer{Barrett and Westbury:
Equals}\refname\Equals.

The path integral theory which is equivalent was described by Witten
\refer{Witten 2+1 gravity}\refname\WittenGrav, and is the theory determined by
the action for three-dimensional gravity with a cosmological constant. With the
appropriate
choices of signature, this action is the difference of two independent
Chern-Simons actions with group $\SU(2)$. This \text{means}, at least on the
heuristic
level of path integrals, that the partition function is the modulus square
of the partition function for the Chern-Simons theory which Witten famously
associated with knot theory \refer{Witten
Jones-polynomial}\refname\WittenJones. This later
theory was defined rigorously by Reshetikhin and Turaev \refer{}\refname{\RT}
by combinatorial methods which were suggested by the properties of Witten's
theory. Therefore the modulus square of the Reshetikhin-Turaev partition
function can be taken to be a definition of three-dimensional gravity with
a cosmological constant. Roberts\refer{}\refname{\Roberts} proved that this
is equal to the Turaev-Viro partition function, thus providing the final
link to interpret the Turaev-Viro theory as a physical model.

The question of the corresponding physical interpretation of the TQFTs based
on other quantum groups remains open.

The limiting procedure referred to above is to take the quantum parameter
$q\to 1$ in the Turaev-Viro theory, or equivalently, the level $k\to\infty$
in the Chern-Simons theory. This corresponds to taking the cosmological
constant $\Lambda\to0$ in the Einstein action. It is not known whether
such limits exist. However, it would be expected that $q\to1$ leads to the
Ponzano-Regge theory on the one hand, and $\Lambda\to0$ to the
three-dimensional gravity on the other hand.

\subhead\nofrills C. Relevance to four dimensions\endsubhead

It is interesting to speculate on whether the experience with three-dimensional
theories tells us anything about four dimensions. In a TQFT,
$$Z(S^1\times\Sigma)=\tr(P_\Sigma)=\dim (\image P_\Sigma),$$
a well-defined number. Therefore the space of `propagating modes' is
\finitedimensional. If, on the other hand, this space is infinite-dimensional,
as is widely believed to be the case for quantum gravity, then
$Z(S^1\times\Sigma)$ is infinite and the theory is not completely definable.
This argument is less clear-cut than it may seem, as we know little about the
space of propagating modes in quantum gravity. When local observables are
included in the theory, as considered in the next section, the theory has a
sequence of similar such spaces of finite but arbitrarily large dimension,
where the dimension increases with the resolution of the measurements. It is
possible that such a construction is an effective substitute for an
infinite-dimensional state space.

The other factor which has to be kept in mind is the different nature of
topology in three and four dimensions. The manifold $S^1\times\Sigma$ has an
infinite fundamental group, which one can conjecture is the reason that its
partition function is not defined in the infinite-dimensional case. Indeed this
is the case in three-dimensional gravity\cite{\WittenTop}. The topology of
closed 3-manifolds is dominated by the fundamental group, as there are no known
simply-connected 3-manifolds other than $S^3$. Therefore the theory is
drastically restricted if it is limited to invariants of simply-connected
3-manifolds or 3-manifolds with finite fundamental group. This is not such a
restriction in four dimensions, where the diversity of the simply-connected
manifolds is the fundamental issue of four-dimensional topology.

The three-dimensional gravity theory exists at all because it is defined for
certain manifolds $M$ with boundary. These are where each element of the
fundamental group is generated by a curve lying in the boundary of $M$. It is
interesting to note that the consistency conditions in section {\smc IV C} and
the example in section {\smc IV D} also led to restrictions of a somewhat
similar nature on the relation of $M$ to its boundary.

\head VI. Local observables\endhead

A desirable feature for a theory of physics is localisation, the
ability to decompose space into subsets, which may have boundary points
in common, to discuss the physics of each subset separately, and
the interaction between them due to the common boundary points. Localisation
is a fundamental property of the standard theories of dynamics, the
theory of fields being a local explanation for action at a distance.

For example, Newtonian gravity was originally formulated as a theory
with an inverse-square law force, but can be reformulated using Poisson's
equation for the potential. This equation can be
solved in a region $A$ of space providing the potential is given on the
boundary of $A$. The boundary data is all that one needs to know about the
sources in the rest of space. If $B$ is a second region of space, disjoint from
$A$ except on the boundary, it is clear that the boundary data for $B$ has to
be equal to the boundary data for $A$ for the points which coincide.

Now consider the case of topological quantum field theory.
Let $\Sigma$ be a closed $(d-1)$-manifold, which can be thought of as `space'.
Consider a decomposition
$$\Sigma=A \cup B,$$
where the manifolds $A$ and $B$ have common boundary $A\cap B$. An observable
which uses $A$ and $B$ as data is necessarily a local observable; as explained
above, a global observable has an action which is invariant under isotopies
of $\Sigma$, and these isotopies do not preserve the decomposition into
$A$ and $B$. An example of such an observable might be
$$\text{There is a particle in $A$.}$$
Let this observable be represented by the projector $\Cal O$ in $V(\Sigma)$.
Since $\Cal O$ is not global, it `sees' the kernel
$$\ker(P_\Sigma)=\{v\colon P_\Sigma(v)=0\},$$
 the part of $V(\Sigma)$ that is redundant from the point of view of the Atiyah
axioms. This is
because $(n-1)$-manifolds with boundary, such as $A$ and $B$, are not
part of Atiyah's scheme.

A concrete example of these constructions can be given in the three-dimensional
state-sum models. The state space for a surface is constructed with the aid of
an arbitrary triangulation, which we can assume is such that $A$ and $B$ are
the union of complete triangles. Then there are state spaces $V(A)$ and $V(B)$,
and
$V(\Sigma)$ is obtained as a certain subspace of $V(A)\otimes V(B)$. This
subspace is determined by a condition that certain boundary data common to $A$
and $B$ are equal. These are the labels on edges.

It is worth contrasting the operator $\Cal O$ with $\Cal G=P_\Sigma\Cal O
P_\Sigma$, which is a global observable, and thus `gauge invariant' in the
terminology borrowed from Yang-Mills theory. If $\Cal O$ were global then these
would be equal. However in the case of a local observable these operators are
not the same. For example, the norm of $\Cal G\C M\ket$ is not necessarily
equal to $\Cal O\C M\ket$, as $\Cal O$ and $P_\Sigma$ do not commute.

The operator $\Cal G$ corresponds to an observable constructed from the
topological data given by identifying the decomposition $A\cup B$ with the
middle $\Sigma\times \{0\}$ of the cylinder $\Sigma\times[-1,1]$.
This is diffeomorphic to the same construction with $A$ and $B$ replaced by
their images under an isotopy of $\Sigma$, $A'$ and $B'$. Therefore the global
observable cannot distinguish between $A$ and $A'$ or $B$ and $B'$. This
accords with the fact that global observables commute with isotopies. The
fact that $\Cal G$ is not a projector is related to this. One cannot prove,
as was done in section {\smc IV B} for global observables, that repeating the
proposition $\Cal O$ at different times, which is the same as repeating $\Cal
G$, is reliable, in the sense that the proposition at one time implies the same
proposition at a later time. Indeed, it does not follow logically that
$$\text{There is a particle in $A \Leftrightarrow$ There is a particle in
$A'$},$$
since $A$ and $A'$ need not coincide.

The logical approach to quantum mechanics requires that observables are
associated with particular sets, such as $A$, and not their equivalence classes
under diffeomorphism or isotopy. This is because the theory is centred on the
logical relation between propositions, which is governed by the relations
between the corresponding sets. A topological relation such as `$A$ intersects
$B$' makes sense whereas there is no such relation between equivalence classes.
The difficulty with the reliability of $\Cal O$ is that the TQFT has not
registered any one particular identification of $\Sigma\times\{t_1\}$ with
$\Sigma\times\{t_2\}$.

For this reason, observables referring to specific regions cannot be gauge
invariant. Gauge invariance is a necessary feature of observables in Yang-Mills
theories, but one has to bear in mind that general relativity is not a
Yang-Mills gauge theory. It may seem puzzling that particular sets are singled
out in the theory, when the
theory is explicitly invariant under the action of diffeomorphisms in an
appropriate way. At one level, this can be seen as a book-keeping device,
the need to name one particular set $A$, so that it can be referred to
explicitly by other related propositions. However there is a somewhat more
physical explanation.

In general relativity, the set of all space-time events is a manifold.
Therefore it is clear that specifying a particular property for the space-time
events specifies a particular subset of this manifold. Diffeomorphisms of the
space-time manifold do not play a basic role in the physical interpretation of
the theory. Even the points of space-time which are `unobserved' are determined
by the unique collection of geometries of paths which lead to them from a fixed
base-point \refer{jwb}\refname\Holonomy. However, when the space-time events
are {\it represented}, either mathematically as points in a coordinate chart,
or say in a computer, then there is an arbitrariness in the representation, in
general, which is accounted for by the action of diffeomophisms.
The simplest and clearest interpretation of TQFT is that these features of
general relativity continue to hold, namely, that the points of the manifold,
perhaps together with some other data, each have an unambiguous physical
meaning. This can be weakened somewhat, in that one does not require an
interpretation for each and every point on the manifold, but for the elements
of the relevant combinatorial presentation of the manifold, which is of a
finite nature. This point reinforces the topological principle which was
described in section {\smc IV C}.

Since diffeomophisms of $\Sigma$ act in $V(\Sigma)$, one can formulate a
condition that the observable $\Cal O$ depends {\it only} on the topology of
$A$
and $B$, in the same way that global observables depend only on the topology of
$\Sigma$.

This is that if $h$ is an isotopy $\Sigma\times [0,1]\to\Sigma\times[0,1]$,
for which the sets $A\times [0,1]$ and $B\times[0,1]$ are invariant (for
example $h(a,t)\in A\times\{t\}$), then
$$h \Cal O=\Cal O h=\Cal O.$$

Suppose $P_\Sigma(A,B)$ is a projector in $V(\Sigma)$ whose image consists of
vectors each fixed by every such isotopy $h$, and whose kernel is invariant by
every such $h$. Then observables which satisfy
$$\Cal O P_\Sigma(A,B)= P_\Sigma(A,B) \Cal O=\Cal O$$
satisfy this criterion. This can be described as saying that they are global
relative to $\{A,B\}$.

If the theory provides such a projector, then it is natural to associate it
with the $d$-manifold $\Sigma\times[0,1]$ equipped with the decomposition
into the two subsets $A\times[0,1]$ and $B\times[0,1]$. Clearly if this
projector is used for `time evolution' instead of $P_\Sigma$, then one can
prove that the observable $\Cal O$ is reliable for measurements repeated in
time. Intuitively, this is because specifying the subset $A\times[0,1]$
provides the required identification of the set $A$ at different times.

The picture which emerges is that the more structure on $\Sigma$ which is
fixed, the larger the state space becomes ($\image P_\Sigma\subset\image
P_\Sigma(A,B)$), and the correspondingly smaller the symmetry group becomes.
The specification of a particular metric tensor, for example, can be regarded
as a limiting case of this procedure. This can be considered in the context of
the Turaev-Viro model. Any given metric tensor can be approximated to any given
accuracy by picking a sufficiently fine triangulation and assigning a length to
each edge \refer{Barrett and Parker}\refname\Parker.

A systematic investigation of the local observables goes beyond the general
framework of topological quantum field theory. The set $A\times[0,1]$ is an
example of a manifold with corners. The boundary is
$$\partial A\times[0,1]=\bigl(A\times\{0,1\}\bigr)\cup \bigl(\partial
A\times[0,1]\bigr),$$
which is a union of $(d-1)$-manifolds with boundary. The boundary of these
is the closed $(d-2)$-manifold $\partial A\times\{0,1\}$, the corners of this
manifold. More generally, one can have corners in all dimensions down to zero,
the basic examples being the cube or the simplex. A set of axioms for
topological field theory which includes manifolds with corners should reduce to
Atiyah's axioms for the special cases where the corners are empty. The notion
of an $n$-category\refer{Street}\refname{\Street} provides a framework for the
general formalism, although at present it is not clear exactly which type of
$n$-category is most useful. In any event, the state-sum models provide
explicit examples of theories with corners.

\enddocument